\documentclass{article}

\usepackage{pbg_template_purple}
\usepackage{pifont}
\usepackage{parskip}

\usepackage[utf8]{inputenc} 
\usepackage[T1]{fontenc}    
\definecolor{MyPurple}{HTML}{6B67EE}
\definecolor{MyPink}{HTML}{ca5670}
\usepackage[dvipsnames]{xcolor}

\makeatletter
\renewcommand\paragraph{\@startsection
  {paragraph}{4}{\z@}
  {\parskip}      
  {-1em}          
  {\normalfont\normalsize\bfseries\color{PennBlue}}}
\makeatother

\usepackage[colorlinks=true, linkcolor=MyPurple, citecolor=MyPurple, urlcolor=MyPurple]{hyperref}
\usepackage[numbers,sort&compress]{natbib}
\usepackage{booktabs}       
\usepackage{amsfonts}       
\usepackage{nicefrac}       
\usepackage{microtype}      
\usepackage{amsmath}
\usepackage{amssymb}
\usepackage{graphicx}
\usepackage{tikz}
\usepackage{subcaption}
\usepackage{colortbl}
\usepackage{algorithm}
\usepackage{subcaption}
\usepackage{tabularx}
\usepackage{amsmath,amssymb,amsthm,bbm}
\usepackage{multirow}
\usepackage{adjustbox}
\usepackage{algorithm}
\usepackage{algpseudocode}
\usepackage{amsmath}
\usepackage[noEnd=false]{algpseudocodex}
\usepackage[toc,page]{appendix} 
\usepackage{tocloft} 
\usepackage{titlesec}
\usepackage{booktabs}
\usepackage{threeparttable}
\usepackage{enumitem}
\setlist[enumerate]{left=10pt}
\usepackage[font=small]{caption}
\usepackage{wrapfig}
\usepackage{listings}
\usepackage{booktabs}
\usepackage{mdframed}
\usepackage{thm-restate}

\usepackage{booktabs}
\usepackage{multirow}
\usepackage{array}
\usepackage{thmtools}
\usepackage{xurl} 
\usepackage{hyperref} 

\title{Rethinking Benchmarks and Models for Enzyme Specificity Prediction}

\author{Elizabeth H Mahood, \textsuperscript{1,2,3, \dag} 
        Natália Komorníková, \textsuperscript{1} 
        Tomáš Pluskal, \textsuperscript{1} 
        Pranam Chatterjee \textsuperscript{3}
    
    \vspace{1em}
    \normalfont \small
    \textsuperscript{1}Institute of Organic Chemistry and Biochemistry of the Czech Academy of Sciences, Flemingovo náměstí 2, 160 00 Prague, Czech Republic\\
    \textsuperscript{2}Computer Science and Artificial Intelligence Laboratory, Massachusetts Institute of Technology, 32 Vassar St, Cambridge, MA 02139\\
    \textsuperscript{3}School of Engineering and Applied Science, University of Pennsylvania, 220 S 33rd St, Philadelphia, PA 19104  
    
    \vspace{0.5em}
    
    \vspace{0.5em}
    \textbf{Correspondence:} \href{mailto:emahood@seas.upenn.edu}{\texttt{emahood@engineering.upenn.edu}} 
}
\begin{document}

\maketitle

\begin{abstract} 
    Artificial Intelligence has had a profound impact on the biological sciences, and in particular has accelerated research on protein form and function. Enzymes — proteins capable of catalysis — are no exception: a surge of predictive models have been recently developed to address a range of enzyme tasks. Models addressing enzyme-substrate or enzyme-reaction compatibility could be especially valuable for enzyme annotation, biosynthetic pathway elucidation, and biocatalyst retrieval — tasks in which a central challenge is the identification of a true catalyst (or truly compatible reaction) among many similar candidates. While existing models report strong performance on alternative benchmarks, less is known about their capabilities in this regime. Herein, we benchmark four recently released enzyme–substrate and enzyme–reaction prediction models, using tasks and datasets tailored to this setting. We first show that two representative enzyme-substrate prediction models perform near random baselines across two enzyme families when considering enzymes and substrates not encountered during training. To evaluate additional models across a consistent dataset, we next assemble the largest cytochrome P450 (CYP) reaction dataset to date—2,922 reactions across 768 enzymes—and construct a CYP ranking benchmark requiring the correct enzyme to be prioritized among all CYPs in its native organism. We again find that most models do not outperform sequence-based (BLAST) baselines even after fine-tuning. We finally adapt the bimolecular structure prediction model Boltz to enzyme-substrate prediction by training supervised classifiers on residue-ligand pair embeddings, and show that this approach consistently surpasses the BLAST baselines on our CYP ranking benchmark. Together, our results argue for more discovery-relevant benchmarking and suggest that interaction-aware representations from full biomolecular complexes may provide a promising basis for enzyme prioritization.
\end{abstract}

\section{Introduction}
Enzymes are highly efficient and specialized catalysts, often capable of performing selective chemical transformations that are difficult or impractical to reproduce using synthetic chemistry alone \citep{devine2018extending}. Their activities are central to metabolism, physiology, and adaptation, making functional annotation an important problem across fields ranging from biocatalyst development \citep{buller2023nature} to ecophysiology \citep{bravo2017arbuscular}. A central component of defining an enzyme’s function is determining the set of reactions it can catalyze, although in practice, the question can often arise in reverse: given a desired reaction, which enzyme is most likely to catalyze it? 

Two broad challenges commonly surface during enzyme characterization \citep{seligmann2024chemical}. In the first, the reaction mechanism is unknown and bears little similarity to reactions catalyzed by previously characterized enzymes, leaving no clear enzyme class or homologous starting point \citep{nett2023plant}. In the second, the reaction is likely catalyzed by a member of a large gene family, yet homology and co-expression offer insufficient resolution to prioritize one enzyme among many plausible candidates, sometimes numbering in the hundreds \citep{caputi2012genome, nelson2011p450}. These candidates often share overall catalytic activity (e.g. methyl transfer) and may exhibit substantial sequence similarity. However, global sequence similarity typically should not be assumed to imply substrate or reaction similarity \citep{mcclune2025discovery, kruse2022orthology}. Indeed, enzymes with high sequence similarity can exhibit distinct substrate preferences \citep{luo2007convergent, weng2013chemodiversity}, while similar reactions may be catalyzed by sequence-divergent enzymes \citep{mcclune2025discovery, bastard2017parallel}. Complementary strategies such as expression profiling \citep{jacobowitz2020exploring} are therefore often used, but these approaches may lack the necessary resolution to isolate positives from co-expressed negatives. These limitations motivate orthogonal approaches that can prioritize enzymes beyond homology alone, particularly in high-candidate settings where only a few enzymes are true matches.

Recent advances in protein machine learning have created new opportunities to tackle enzyme prioritization. Large pre-trained protein language models such as ESM \citep{lin2023evolutionary} and structure-prediction models such as AlphaFold \citep{jumper2021highly, abramson2024accurate} enable enzymes to be represented using learned sequence embeddings and predicted structural features which capture evolutionary and functional information \citep{tran2023survey}. These representations are now commonly combined with molecular graphs, fingerprints, or reaction encodings in supervised models trained across diverse enzymes and chemistries. Broadly, such approaches can be divided into substrate-prediction models, which estimate enzyme–substrate compatibility \citep{kroll2023general, du2025fusionesp, cui2025enzyme, campbell2024viper, qian2024deep}, and reaction-prediction models, which associate enzymes with the reactions they are likely to catalyze \citep{mikhael2024clipzyme, hua2024reactzyme, rocks2025dual, liu2026geometric}.

Since the introduction of ESP \citep{kroll2023general}, the first general enzyme–substrate prediction model to our knowledge, the field has expanded rapidly to include multiple substrate-prediction and reaction-prediction architectures. These models have generally been evaluated in settings that emphasize either global discrimination performance, such as micro-averaged Area Under the Receiver Operating Characteristic (AUROC) and Area Under the Precision-Recall (AUPR) metrics \citep{kroll2023general, du2025fusionesp, cui2025enzyme, qian2024deep}, or retrieval of the correct enzyme or reaction among dissimilar candidates \citep{mikhael2024clipzyme, hua2024reactzyme}. Such evaluations and benchmarks \citep{yang2024care, hua2024reactzyme} are valuable, but they largely probe generalization in a low-homology regime. They do not directly address the high-homology setting, in which the central challenge is to prioritize a few true enzymes from many closely related candidates within an expanded gene family. This gap is particularly important because the latter setting is a common bottleneck in practical enzyme elucidation and pathway discovery \citep{seligmann2024chemical}.

Herein, we sought to evaluate the utility of current substrate and reaction prediction models within this high-homology setting. To this end, we chose two representative substrate prediction models (FusionESP \citep{du2025fusionesp} and EZSpecificity \citep{cui2025enzyme}) and two reaction prediction models (CLIPZyme \citep{mikhael2024clipzyme} and EnzymeCAGE \citep{liu2026geometric}), and quantified their performance on highly imbalanced data, wherein negatives sometimes drastically outweighed positives. To thoroughly test generalizability of these models, we measured their predictive ability on datasets in which both enzymes and substrates/reactions were unseen during training, and quantify their performance using stringent ranking metrics which measure the models’ ability to highly prioritize real substrates/reactions over negatives. Finally, we ask whether representations from co-folding structure-prediction models, exemplified by Boltz \citep{wohlwend2025boltz}, can provide additional information for enzyme–substrate discrimination. Together, these results have implications both for interpreting the reported performance of current enzyme prediction models and for guiding their design in the future.

\section{Results}

\subsection{Representative substrate prediction models show limited generalization abilities}

As an initial assessment, we sought to determine how well two representative substrate prediction models, FusionESP and EZSpecificity, could prioritize substrates for enzymes across different classes. Enzyme screens offer accessible datasets with which to evaluate these models; however, high-throughput enzyme assays typically lack definitive product information, excluding reaction prediction models from this analysis. While several enzyme screens have been collected and processed for this purpose \citep{goldman2022machine}, many of these same screens were included in EZSpecificity’s training data, making it impossible to determine its ability to generalize to new enzyme/substrate pairs. To facilitate this, we identified a Nitrilase (NTL) screen \citep{vergne2013nitrilase} and an \(\alpha\)-ketoglutarate (\(\alpha\)-KG)/Fe(ii)-dependent enzyme screen \citep{paton2025connecting}, neither of which were used by either model for training or evaluation. We processed and binarized activity measurements following the thresholding regime set forth by Goldman et al. For each model, we removed enzymes and substrates present in its training data, and further excluded any remaining enzyme with no recorded positive substrates, since ranking metrics are undefined or uninformative with no active substrates present. For FusionESP, 119 NTLs and 79 \(\alpha\)-KGs remained after filtering, assayed against 24 and 71 substrates, respectively. For EZSpecificity, these enzyme/substrate numbers were 77 and 10 (NTLs), and 75 and 66 (\(\alpha\)-KGs). Note that direct comparisons of model performance are impossible as, after training data subtraction, the final evaluation datasets differed between models.

\begin{figure}
    \includegraphics[width=\textwidth]{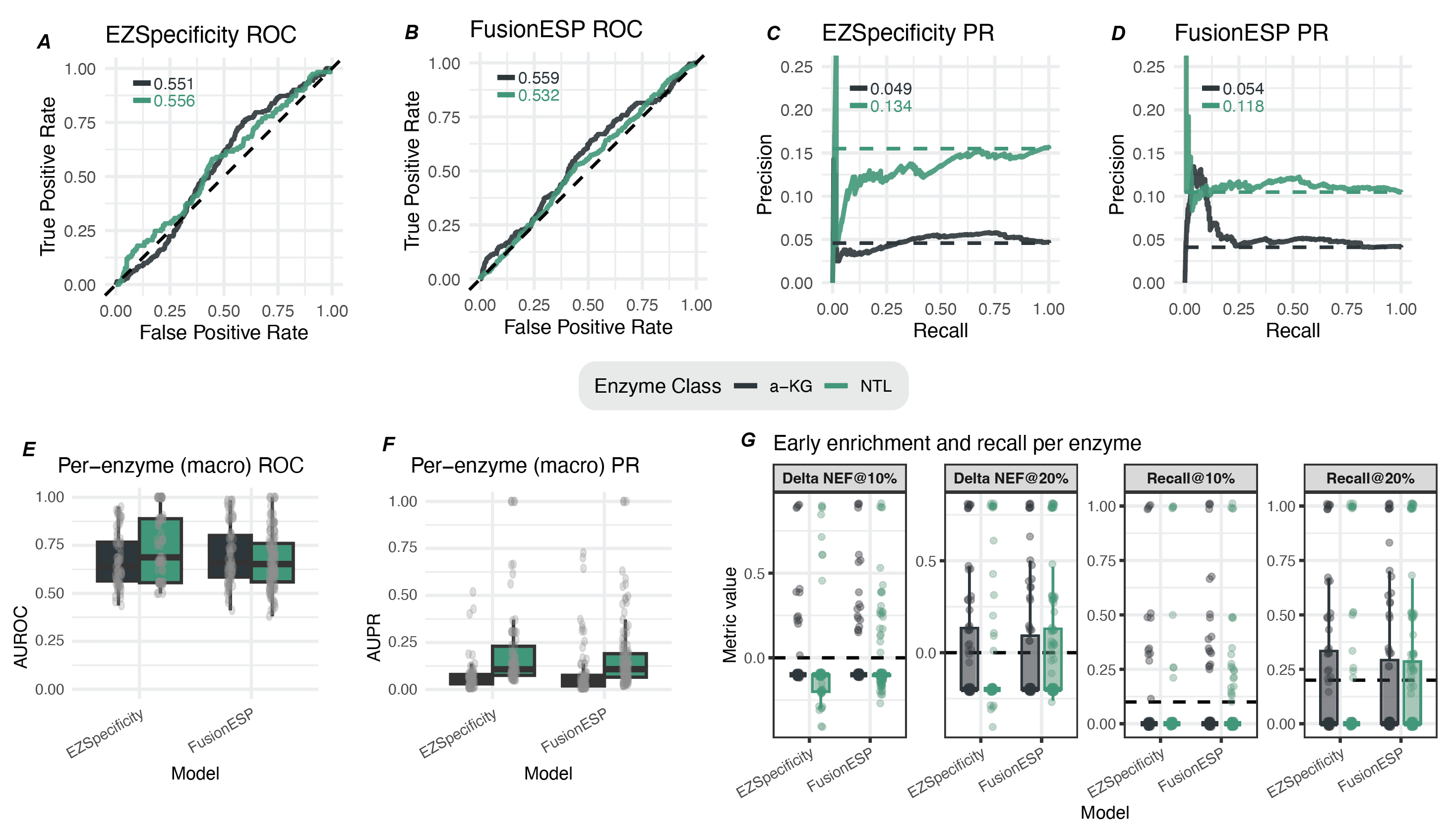}
    \caption{\textbf{Benchmarking two Enzyme/Substrate Prediction models across unencountered enzymes and molecules.} \textbf{(A-D)} Per-model, micro-averaged ROC and PR plots. Dashed lines show classification-at-random baselines. \textbf{(E-F)} Macro-averaged ROC and PR plots. Gray dots represent individual enzyme scores. \textbf{(G)} Recall at 10\% and 20\%, and \(\Delta\)Normalized Enrichment Factor (NEF) at 10\% and 20\% per-enzyme (see Methods for formulation). Colored dots represent individual enzyme scores while dashed black lines represent random baselines. The legend in the middle applies to each subplot. \(\alpha\)-KG = \(\alpha\)-ketoglutarate Fe(ii)-dependent enzymes; NTL = Nitrilases. Note that direct comparisons of model performance are impossible as, after training data subtraction, the final evaluation datasets differed between models.}
    \label{fig:F1-main}
\end{figure}

We first discerned each model’s ability to rank positive enzyme-substrate pairs over negatives using micro AUROC and AUPR (\textbf{Figure \ref{fig:F1-main}, A-D}), metrics which are commonly reported across substrate and reaction prediction models. Across both enzyme classes, both models, and both metrics, the scores obtained were close to random baselines. These values are lower than those originally reported for the models, indicating that both models show limited generalization to the held-out data present in these screens. 

Next, we decomposed the global, pooled AUROC and AUPR metrics into their per-enzyme equivalents (\textbf{Figure \ref{fig:F1-main}, E, F}). This analysis assesses whether, for a given enzyme, the model ranks positive substrates above negative substrates. For macro AUROC, we see most enzymes achieving a score above the random baseline of 0.5, with mean AUROCs ranging from 0.67 (FusionESP on NTLs) to 0.74 (EZSpecificity on NTLs; \textbf{Supplementary Table \ref{tab:benchmark_results}}). Although both models struggle to score positive enzyme-substrate pairs globally higher than negatives, they show better discernment when considering each enzyme individually. This suggests these models may have learned local within-enzyme signal rather than a robust, global “interaction” score consistent across enzymes or enzyme classes. However, most enzymes achieve low AUPR, with only 40\% (EZSpecificity on NTLs) to 60\% (FusionESP on  NTLs) of enzymes beating their random baselines, suggesting a restricted ability to highly rank an enzyme's positive substrates (\textbf{Supplementary Table \ref{tab:benchmark_results}}). 

To more directly determine the models’ ability to rank positive substrates highly among all candidates, we next calculated per-enzyme precision at 10\% and 20\%, as well as \(\Delta\)Normalized Enrichment Factor (\(\Delta\)NEF) values at 10\% and 20\%. These metrics are especially relevant for any end users as they more directly capture how many interactions need screening before a positive is reached. The NEF, a metric borrowed from drug discovery scenarios \citep{caba2024comprehensive}, derives from the Enrichment Factor (EF), which measures the enrichment of positives among the top-k ranked predictions compared to random guessing, with a value of one representing no enrichment, or a random baseline (see \textbf{Methods}). The NEF scales the EF to [0, 1] by dividing by the theoretical maximum EF—the EF achieved if all true positives are ranked higher than true negatives. While normalization allows for comparison amongst enzymes with differing ratios of positives, especially those below 10\% to those well above, it can obscure comparisons to a random baseline. To explicitly show the model’s performance relative to a random classifier, we report the \(\Delta\)NEF, or the difference between the realized NEF and a random baseline (defined as 1 / the theoretical maximum EF), for each enzyme.

We again find a minority (from 10\% for EZSpecificity on NTLs to 23\% for FusionESP on \(\alpha\)-KGs) of enzymes that have above-random early enrichment of positive substrates among the top 10\% of predictions (\textbf{Figure \ref{fig:F1-main}, G; Supplementary Table \ref{tab:dnef_results}}). These numbers increase to between 20\% (EZSpecificity on NTLs) and 32\% (EZSpecificity on \(\alpha\)-KGs) as k increases to include the top 20\%, suggesting ranking capabilities of both models are moderate rather than high. When considering global rather than per-enzyme Recall@10\% (\textbf{Supplementary Table \ref{tab:benchmark_results}}), scores are universally low, ranging from 0.07 (EZSpecificity on NTLs) to 0.15 (FusionESP on \(\alpha\)-KGs), further indicating that these models are unlikely to have learned a generalizable, robust proxy for enzyme-substrate interaction. While these values increase when considering Recall@20\% — from 0.15 for EZSpecificity on NTLs to 0.23 for FusionESP on \(\alpha\)-KGs — they are not appreciably better than the random baseline. 

\subsection{Creation of a CYP dataset for task-specific learning}
We next sought to evaluate an extended set of models, specifically including reaction prediction models. We further wished to curate a class-specific dataset which would allow us to (i) determine the extent to which fine-tuning could increase task-specific performance, and (ii) create a controlled, leakage-aware split which would enable fair comparisons among models. To this end, we assembled the largest dataset of CYP sequences and associated reaction information (substrate and product SMILES) to date. We focus this analysis on CYPs because they are ubiquitous oxygenases, prevalent across many biosynthetic pathways and organisms, and their substrates are notoriously difficult to predict through current, sequence-based approaches. While there are pre-existing databases of CYP-catalyzed reactions \citep{zhang2024p450rdb, wang2021pcpd}, they are taxonomically limited and small (containing on the order of dozens to hundreds of reactions). We instead chose to gather as much information as possible on CYP reactions across different organisms, substrates, subfamilies, and (known) substrate counts. 

\begin{figure}
    \centering
    \includegraphics[width=\textwidth]{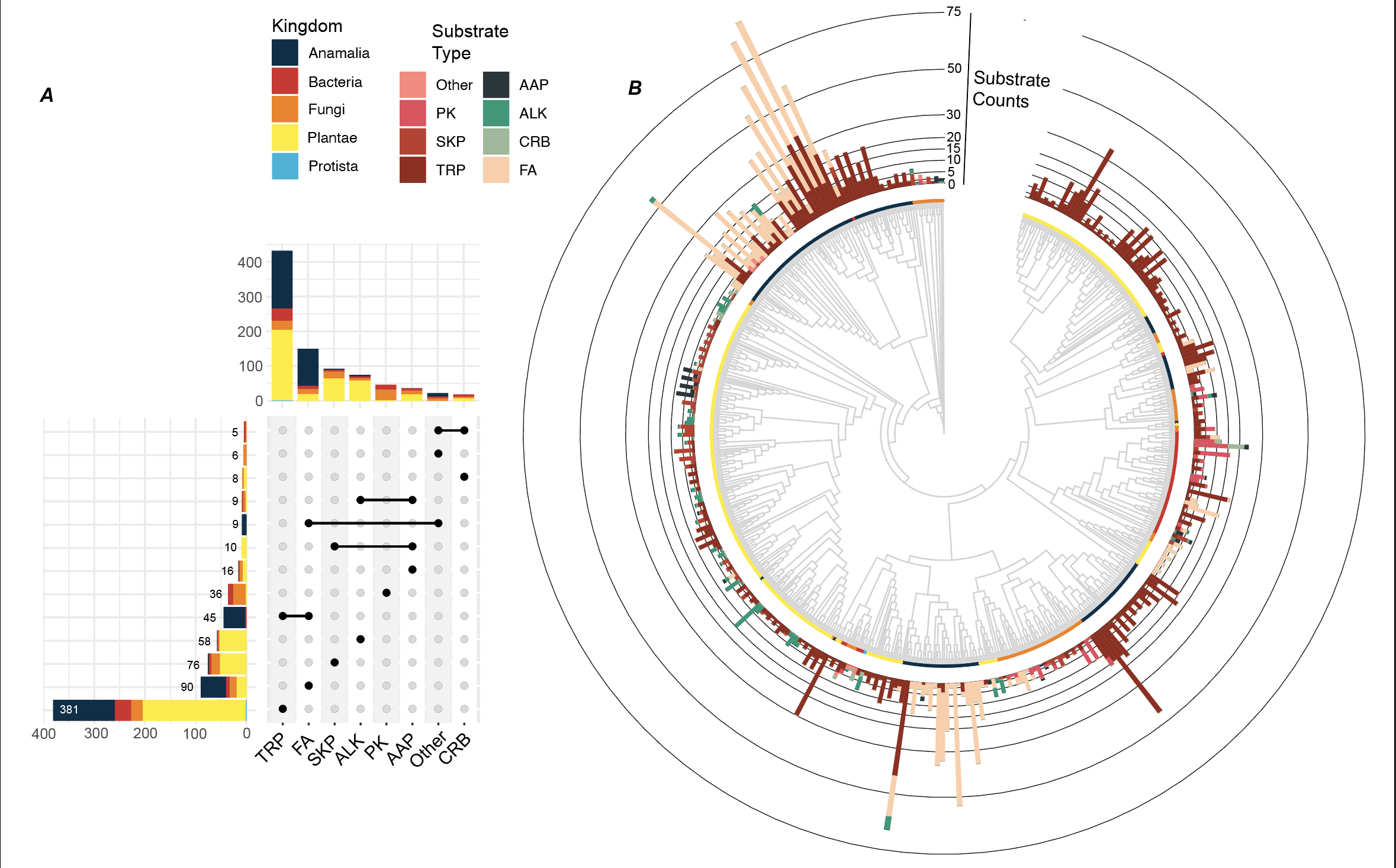}
    \caption{\textbf{Substrate utilization among the collected CYP enzymes.}  \textbf{(A)} Substrate usage patterns across compound class and kingdoms. The top barchart maps total substrate counts of each compound class, distributed across kingdoms. The side barchart charts the number of enzymes utilizing substrates across different combinations of compound classes. Only combinations with at least five enzymes are plotted. \textbf{(B)} Phylogenetic tree depicting enzyme kingdom (inner plot) and substrate counts (outer plot) for collected CYP enzymes. Substrate counts are summed across every two enzyme neighbors. Compound types: AAP = Amino acids and Peptides; ALK = Alkaloids; CRB = Carbohydrates; FA = Fatty acids; PK = Polyketides; SKP = Shikamates and Phenylpropanoids; TRP = Terpenes. The Kingdom legend applies to both A and B subfigures.}
    \label{fig:F2-cyp}
\end{figure}

To this end, we sourced CYP-catalyzed reactions from existing CYP databases \citep{zhang2024p450rdb}, and augmented them with CYPs sourced from UniProt. We limited inclusion to manually annotated enzymes with sequences matching PFAM domain PF00067, and with full substrate and product information recoverable from RHEA. In total, this process yielded 2922 reactions across 768 proteins and 616 substrates. The included enzymes utilize substrates spanning major classes of natural products (\textbf{Figure \ref{fig:F2-cyp} A, B}), and are derived from all kingdoms except Archaea. We note that, while we originally collected entries across different reaction chemistries, including cyclization and decarboxylation reactions, we later reduced the dataset to contain only mono-oxygenation reactions to facilitate data processing and model training. 

With our dataset assembled, we next processed it further into a format amenable for training and evaluating machine learning models. A primary goal was to create an evaluation dataset that mimics the process of enzyme elucidation while accurately testing the generalization capabilities of the model. As the first step in creating the evaluation dataset, we isolated 45 reactions with enzymes and substrates disjoint ( < 40\% sequence identity and < 0.7 Tanimoto similarity) from any reaction in the remainder of the data. For each reaction, we then recovered all putative CYP enzymes from the species of origin, and measured the rank at which the correct enzyme was placed amongst all CYPs. We note that, due to the promiscuous capabilities of CYPs, we do not expect the correct enzyme to be consistently ranked at the very top, and rather evaluate the models’ performance at ranking the correct enzyme within the top 10, 20 or 50\% of enzymes. Further, we ensure that none of the nine highly promiscuous, detoxifying human CYP enzymes \citep{nebert2002clinical} are included in the evaluation dataset, while noting that recent evidence implicates many biosynthetic plant and fungal CYPs as being more promiscuous than initially thought \citep{werck2023promiscuity} – indeed, a majority of the CYPs retrieved from databases (423/768, 55\%) catalyze at least two substrates. We deem this evaluation dataset the \textbf{\textit{CYP ranking}} dataset. The remainder of the dataset, henceforth referred to as the \textbf{\textit{CYP training}} dataset, was used for model training, while reserving 25\% of reactions for the validation dataset based on enzyme sequence similarity (< 40\%). 

\subsection{Model performance on the CYP ranking dataset}
We next evaluated the performance of two substrate prediction models (FusionESP and EZSpecificity) and two reaction prediction models (CLIPZyme and EnzymeCAGE) on the CYP ranking dataset. We chose these four models to represent the current state-of-the-art while being inclusive of different architectural choices. While each model incorporates ESM embeddings into enzyme featurization, all models except FusionESP additionally include enzyme structure. CLIPZyme further utilizes EC class to rank candidate enzymes from most to least likely to perform a candidate reaction. EZSpecificity and EnzymeCAGE consider the structure of the entire enzyme-substrate complex, and filter the enzyme representation to retain only residues under a certain distance away from the substrate. While any differences in performance cannot be exclusively attributed to these highlighted architecture choices, this diversity could allow us to assess how differing architectural decisions influence performance on our CYP ranking task.

We evaluate these models both as-is, as well as after retraining their architectures on the CYP training dataset. This process mimics “fine-tuning”, in which an existing general-purpose model is adapted for a specialized task, typically leading to performance gains. Further, retraining all models on the same data enables a more direct comparison of model performance, removing the impact of data similarities between the model's original training data and our ranking data. Our metrics include: the percentage of reactions for which the correct enzyme was ranked in the top 10, 20, and 50\% of enzymes, as well as the EF at 5\%, the Boltzmann-Enhanced Discrimination of the Receiver Operating Characteristic (BEDROC) with \(\alpha\) set to 20 (emphasizing the top ~5\% of predictions), the mean percentile and the Mean Reciprocal Rank (MRR) – which records how high, on average, the model placed the correct enzyme across all 45 reactions. As baselines, we include a random classifier, as well as two proxies for sequence or homology-based enzyme discovery methods. The first, BLAST-sequence, scores and ranks each candidate enzyme by its maximum sequence similarity to any enzyme in the CYP training dataset. The second, BLAST-substrate, first finds the substrate in the CYP training dataset with the highest Tanimoto similarity to the reaction substrate, and scores and ranks all candidate enzymes by their pairwise similarity to the enzyme(s) accepting that substrate.

\begin{figure}
    \centering
    \includegraphics[width=\textwidth]{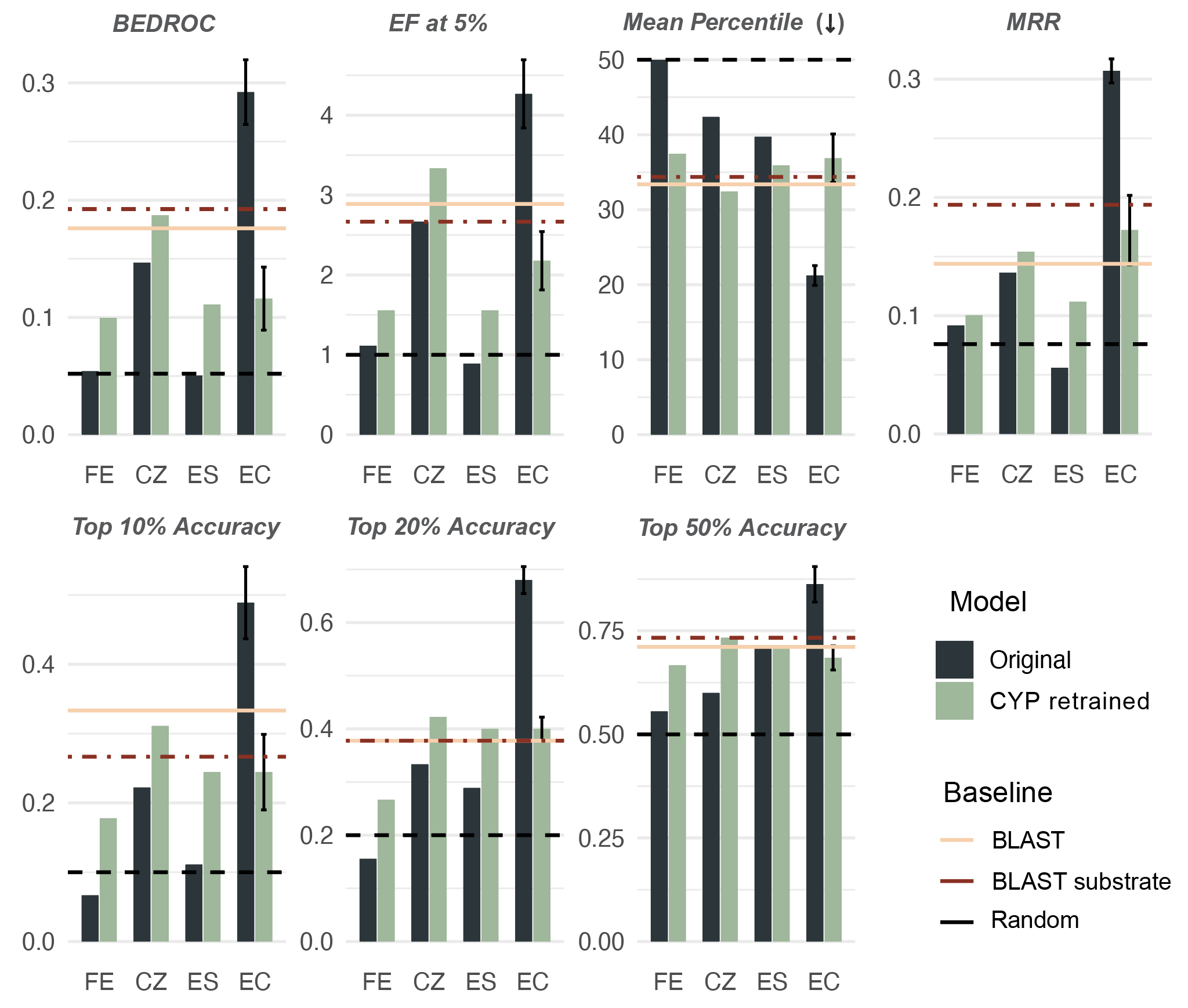}
    \caption{\textbf{Performance of FusionESP, CLIPZyme, EZSpecificity, and EnzymeCAGE against BLAST and Random baselines on the CYP evaluation dataset.}  Higher scores are better for all metrics excepting Mean Percentile. BEDROC = Boltzmann-Enhanced Discrimination of the Receiver Operating Characteristic (alpha at 20); EF = Enrichment Factor; MRR = Mean Reciprocal Rank; FE = FusionESP; CZ = CLIPZyme, ES = EZSpecificity, EC = EnzymeCAGE. Both the BEDROC and Enrichment Factor are calibrated to score the enrichment of rankings of the correct enzyme among the top 5\%. We evaluated five EnzymeCAGE model seeds. The mean score for each metric is plotted as the bar height; error bars represent standard deviation.}
    \label{fig:F3-submods}
\end{figure}

We first consider FusionESP and CLIPZyme, models which do not rely on the structure of the substrate-enzyme complex for featurization. As anticipated by the evaluation of FusionESP in Section 1, we again find that FusionESP never appreciably outperforms the random baseline (\textbf{Figure \ref{fig:F3-submods}}). CLIPZyme, however, consistently outperformed  both random baselines and FusionESP. As expected, retraining on the CYP training dataset improves prediction performance for both models. Taken together, these results could suggest that the enzyme features dictating substrate scope, which are in general not conserved and oftentimes even under positive selection \citep{dueholm2015evolution, almeida2016whole}, are not highly weighted when models are trained using the complete enzyme representation and data collected across diverse enzymes or substrates/reactions. Instead, these models may assign more importance to larger, more conserved features which distinguish one enzyme class from another. However, when considering only one class of enzymes, as is the case for the CYP-retrained models, any global features common across all enzymes (e.g. the heme-binding domain in CYPs) lose discriminatory power, placing more saliency or “importance” on the features dictating substrate scope. 

We next consider EZSpecificity and EnzymeCAGE. While these models have similar methods of feature extraction, they achieve a striking difference in performance. EZSpecificity achieves modest improvements over random baselines only after fine-tuning, while the original version of EnzymeCAGE is the only model to consistently exceed BLAST baselines. Surprisingly, after retraining (\textbf{Figure \ref{fig:F3-submods}}; EC-CYP retrained bars) EnzymeCAGE’s performance consistently declined. This initially unexpected result prompted us to examine the relationship between our ranking dataset and EnzymeCAGE’s original training data. As both were sourced from RHEA \citep{bansal2022rhea}, we suspected that overlap between the two datasets could be inflating apparent performance. Indeed, for 39 of 45 ranking reactions, either the true substrate or the true enzyme was present in EnzymeCAGE’s training data. In light of this substantial overlap, performance after fine-tuning may provide a more informative estimate of generalizability to unseen substrates and enzymes. \textbf{Supplementary Figure \ref{fig:S_EC_pruned}} shows EnzymeCAGE’s performance on the six remaining reactions. Taken together, these results show that all models are capable of strong performance increases over random baselines when trained on task-specific data. However, EnzymeCAGE is the only model to consistently exceed the BLAST baselines -- and this advantage should be interpreted cautiously since it may be partially confounded by training–evaluation overlap.

\subsection{CYP ranking using the bimolecular-complex structural prediction tool Boltz}
We theorized that co-opting a model initially trained for structure prediction of biomolecular complexes may yield competitive performance when predicting enzyme-substrate interactions. Mounting evidence points to the capacity of these models to obey physical and chemical constraints, presumably “learned” during the process of recapitulating the placements of intermolecular interactions \citep{roney2022state}. Further, binding affinity prediction models built upon structure prediction models have shown promising performance \citep{lai2024interformer, passaro2025boltz}, providing further motivation to apply this technique to enzyme-substrate interaction prediction.

\begin{figure}
    \centering
    \includegraphics[width=\textwidth]{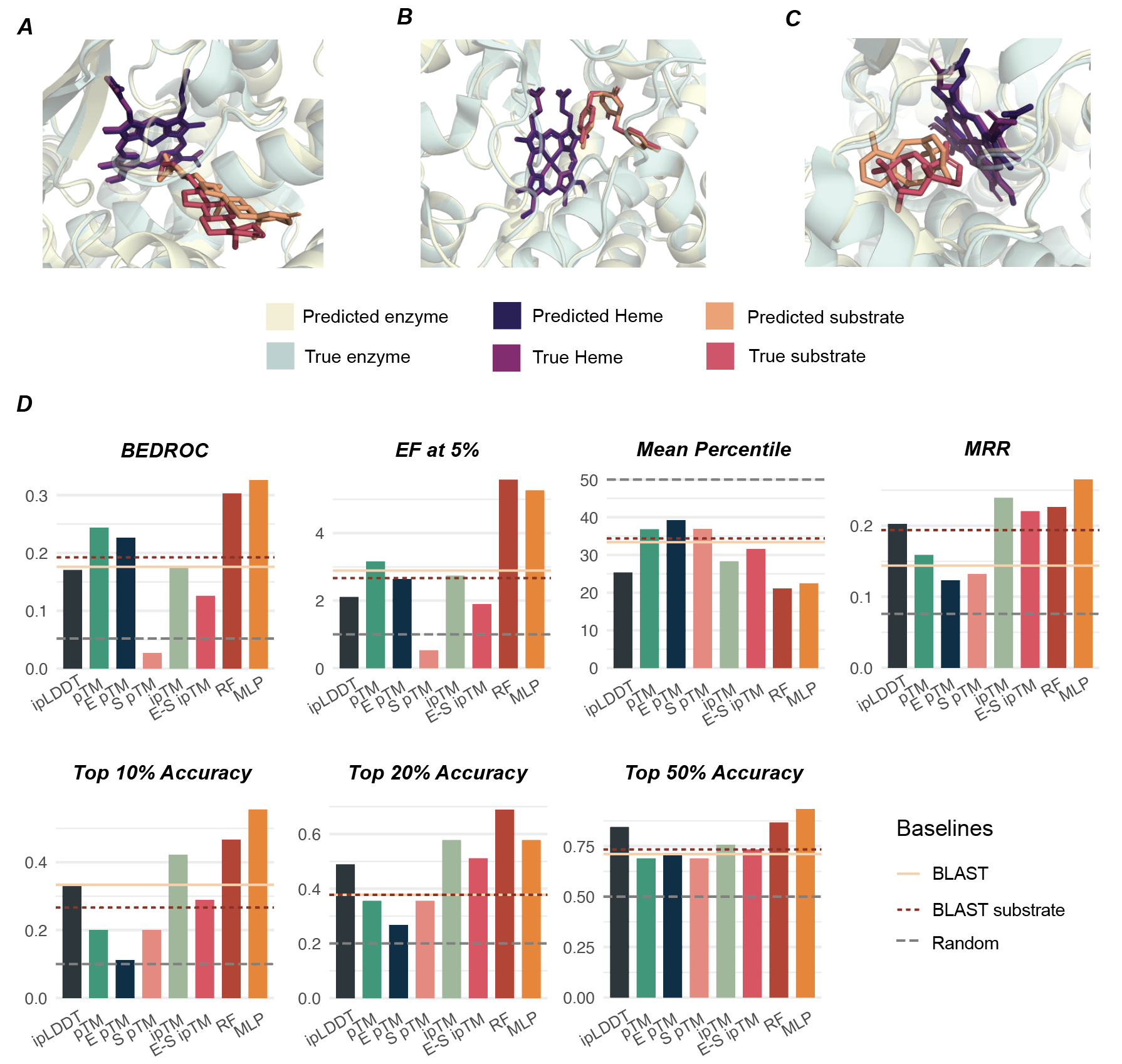}
    \caption{\textbf{Assessment of Boltz’s CYP Complex Structure Prediction and Performance on the CYP Evaluation Dataset.} \textbf{(A)} Structure complex (enzyme, heme, substrate) predicted by Boltz overlaid with PDB structure complex for 8A6W. \textbf{(B)} 8TDQ. \textbf{(C)} 8X3E. \textbf{(D)} Performance of Boltz confidence metrics and the RF and MLP models against BLAST and Random baselines on the CYP evaluation dataset. Higher scores are better for all metrics except Mean Percentile. BEDROC = Boltzmann-Enhanced Discrimination of the Receiver Operating Characteristic (alpha at 20); EF = Enrichment Factor; MRR = Mean Reciprocal Rank; ipLDDT = interface predicted Local Distance Difference Test; (i)pTM = (interface) predicted Template Modeling; E = Enzyme; S = substrate; E-S = Enzyme-Substrate; RF = Random Forest; MLP = Multilayer Perceptron.}
    \label{fig:F4-boltz}
\end{figure}

To de-risk this approach, we first assessed the quality of Boltz’s predictions on CYP-heme-substrate complexes. We obtained the structures of the three complexes in the PDB that contained CYP enzymes and their native substrates, yet were not present in Boltz’s training data (deposit date after June 1, 2023). Although the global complex accuracy was high (< 0.61 Å), we were particularly interested in the positioning of the substrate relative to the heme, as well as the active site residues. Although we were unable to quantify the RMSD error in substrate placement due to atom naming mismatches, we find accurate positioning of the Heme cofactor and the six substrate recognition sites defining the active site (\textbf{Supplementary Table \ref{tab:S_boltz_rmsd}}) for two of the three complexes. Qualitatively, we observe consistent recreation of the substrate’s orientation within the active site (\textbf{Figure \ref{fig:F4-boltz} A, B, C}) and specifically towards the catalytic Fe in the Heme, in these same two complexes. In the third complex, CYP725A4 from \textit{Taxus cuspidata} containing taxadiene, the substrate’s predicted orientation is opposite to the true. We also find that, while the placement of the Heme and SRSs were globally correct (\textbf{Supplementary Figures \ref{fig:S_8a6w}, \ref{fig:S_8tdq}, \ref{fig:S_8x3e}}), the predicted orientation of the Heme was rotated 180 plane-wise. From this initial analysis, we concluded that although the active site components were faithfully grouped together, accurate substrate placement was not guaranteed. Thus, rather than build a substrate prediction model using features derived solely from the predicted structural complexes, we chose to instead utilize the model’s learned internal representations of ligand-residue interactions. Similar to how large language models derive information-dense representations of individual words through extensive training, modern structural prediction models learn these representations for individual atoms, as well as interactions between pairs of atoms in the complex. We utilize these representations following the methods of Passaro et al. \citep{passaro2025boltz}, whereby we pull interaction (or “contact”) representations from all ligand atom-protein residue pairs in which the residue is < 5Å from any atom in the substrate or heme (see \textbf{Methods}). 

We train two output predictors (or “heads”), one a multi-layer perceptron (MLP) and the other a random forest (RF), to discriminate positive from negative CYP-substrate pairs using these embeddings. We additionally evaluate the metrics and scores generated by Boltz’s confidence module, which output the probability of the global and/or local structural predictions being correct (\textbf{Figure \ref{fig:F4-boltz} D}). In theory, such metrics could capture the likelihood with which a given substrate is to be found within a particular enzyme’s active site. We find clear though inconsistent examples of these scores exceeding the BLAST baselines, notably the iPTM score’s Mean Reciprocal Rank, and accuracy at 10 and 20\%. However, iPTM scores lower than the BLAST baselines on the BEDROC and EF at 5\%, indicating its ability to discern and rank the correct enzyme is moderate rather than strong. The consistent out-performance of both BLAST baselines is demonstrated only by the output heads, trained in a supervised manner on CYP-specific data, and rivaling the most performant enzyme-substrate model. 

We conclude from these results that Boltz, and in theory all generative models trained to represent biomolecular structural complexes, can create embeddings that assist downstream models in predictive or discriminative tasks. Herein, we show that such a model outperforms existing models (FusionESP, CLIPZyme, EZSpecificity) on a difficult task – discerning the correct (positive) enzyme-substrate pairs out of highly similar negative pairs. Importantly, we demonstrate that several current models designed and optimized to place substrates with their catalyzing enzymes, across broad substrate and enzyme classes, are not likely to outperform BLAST at this task. 

\section{Discussion}
In this work, we benchmarked several recent enzyme–substrate and enzyme–reaction prediction models under ranking settings that reflect practical enzyme discovery. We first found that FusionESP and EZSpecificity, despite previously reported discrimination performance, performed near random baselines when tasked with ranking true substrates among many negatives across two enzyme families. We then constructed a CYP-specific benchmark for broader comparison and observed that most current models did not consistently surpass random or sequence-based BLAST baselines, even after fine-tuning. Of the specialized prediction models, EnzymeCAGE showed the strongest overall performance, although the significant overlap between training and evaluation data may inflate apparent generalization performance. A discriminative approach built on Boltz pair representations was the only method to consistently outperform the BLAST baselines on a leakage-aware dataset. These results have implications both for how current models are evaluated and interpreted, and for the design of future enzyme-prioritization models.

The performance of substrate/reaction prediction models demonstrated herein is in contrast to the superlative results (AUROCs, AUPRs and/or accuracy often exceeding 0.9) originally reported by these models. Our results suggest that benchmarking practices for enzyme–substrate and enzyme–reaction prediction should be aligned more closely with practical enzyme discovery. Common training schemes for enzyme-substrate models \citep{kroll2023general, du2025fusionesp, qian2024deep, cui2025enzyme} often pair each positive example against only a few negatives, whereas real discovery settings typically require prioritizing a small number of true hits from much larger candidate pools. Similarly, although global discrimination metrics such as AUROC, AUPR, and F1 remain useful summary statistics, they do not directly quantify early retrieval performance or experimental screening burden. For discovery-oriented applications, we therefore suggest greater emphasis on ranking-based metrics such as enrichment scores, BEDROC, MRR, and top-K accuracy. These metrics, which are common in the reaction ranking models (initially introduced by CLIPZyme) and drug discovery settings \citep{caba2024comprehensive, boldini2024machine}, yet are still rare among substrate prediction models, are a more direct measurement of how many enzymes would need to be tested before finding a positive hit. We further suggest that train–test partitioning should consistently enforce meaningful dissimilarity for both enzymes and substrates. In more general studies on protein-ligand interactions, such ”difficult” splits are the standard, and more permissive splits have been shown to be capable of substantial leakage \citep{corso2024deep, durairaj2024plinder}. Indeed, some enzyme-substrate models show strong performance decreases when moving to substrates or enzymes not included in the training data \citep{kroll2023general, cui2025enzyme, atabaigi2026information}. Evaluating on harder splits, together with more stringent metrics, could make reported performance more informative for the settings in which these models are most likely to be deployed.

We find that the Boltz-heads are consistently able to surpass the BLAST baselines, and speculate that this ability may be attributable to two important factors. First, we represent enzyme-substrate pairs jointly via embeddings from a co-folding structure prediction model, rather than train a dual-encoder model on separate molecule and enzyme representations. Since the release of AF2 \citep{jumper2021highly}, a growing body of work has argued that structure prediction models have “learned” biophysical laws governing protein folding \citep{gut2025dissecting, alquraishi2021differentiable, roney2022state}, and can adhere to steric and chemical constraints as a consequence. This evidence engenders the exciting, though speculative, possibility that embeddings from co-folding models may contain information specific to the interactions between enzyme and substrate — information that protein-only or molecule-only embeddings could never capture. Although the true generalization capacity of these models is still an open question \citep{masters2025investigating, chakravarty2024alphafold, outeiral2022current}, our results indicate that any physical laws or chemical intuition "learned" during training can be distilled and harnessed in downstream discriminative tasks. The second feature of our approach which we believe may contribute to model performance is simply that we do not consider all residues in the enzyme. The substrate scope of an enzyme is often determined by a select few residues (sometimes even just one \citep{morris2020single, tsuda2014single}) within the active site or substrate recognition/entry sites. Due to the sparsity of these key discriminative residues, we argue that using an enzyme encoder (be it ESM or a GNN) that is inclusive of the entire enzyme is likely to minimize or stifle their signal. Counteracting this, however, is CLIPZyme's typically superior performance on the CYP ranking dataset (after retraining), over retrained EZSpecificity's or EnzymeCAGE's, given that CLIPZyme encodes the entire protein whereas EZSpecificity and EnzymeCAGE only encode substrate-proximal residues. However, CLIPZyme's outperformance may be attributed to other unique features of its architecture (e.g. its uinque reaction representation), and the performance benefit of selective representation remains to be ablated.

We conclude by noting several limitations of the present study and directions for future work. Most critically, while the strongest-performing approaches on our CYP ranking dataset exceeded sequence-based baselines, substantial performance gains will still be needed before such models are likely to become routine tools for experimental discovery. In particular, accuracy at the top 10\% of ranked candidates remains well below the level that would provide utility in settings involving hundreds of enzymes. Further, model deployment also remains challenging. The Boltz-head approach is particularly intensive, requiring GPU access and potentially terabytes of storage space. Its applicability might not extend to all enzyme classes, especially any transferase, or ones in which the reactants or reaction mechanisms are difficult to determine (e.g. Terpene Synthases). There is likewise ample room to explore alternative downstream architectures, splitting strategies (including structure-based splits \citep{bushuiev2024revealing}) and training strategies (including confirmation-aware generative models \citep{cavanagh2026computational}); we summarize these potential future directions in \textbf{Supplementary Figure \ref{fig:S_future}}. In our setting, more expressive models such as Transformers and GNNs did not improve performance, likely because of limited training data, and alternative negative-sampling schemes similarly did not outperform simple enzyme–substrate mismatches. We nevertheless include our code for implementing these alternatives strategies and hope that the datasets, code, and evaluation framework introduced here will support further development of enzyme-prioritization models and contribute to improved pathway elucidation and biocatalyst discovery.

\section{Methods}
\subsection{Benchmarking EZSpecificity and FusionESP}
FusionESP's training data was accessed from Zenodo (\url{https://zenodo.org/records/13891018}), and EZSpecificity's training data from Google drive (\url{https://drive.google.com/drive/folders/1CvEiGhMJ11Rid8UWwunwK2OnupolZ02K}). \(\alpha\)-KG screening data was sourced from HuggingFace (\url{https://huggingface.co/gomesgroup/catnip/tree/main/raw\_data}), and Nitrilase data from the Supplemental Figure 1 of \citep{vergne2013nitrilase}. Both screens were binarized and added to the collection of enzyme screening sets found at \url{https://github.com/lizmahood/enzyme-datasets}, with original datasets sourced by \citep{goldman2022machine}. For each model independently, we removed substrates and enzymes from the screening datasets if they overlapped with the training data (more details can be found below). We subsequently deleted any enzyme for which all true positives or negatives were removed. After obtaining model predictions for all remaining enzyme-substrate pairs, we calculated AUROC, AUPR (both micro- and macro-averaged), EF \citep{halgren2004glide}, Normalized EF (NEF), and Recall@k. 

To calculate EF@10\%, substrates for each enzyme were first ranked by predicted score. The enrichment factor was then computed as

\[
EF_{10\%} = \frac{h_{\text{top-}k}}{0.1\,n_{+}},
\]

where \(h_{\text{top-}k}\) is the number of true positives among the top 10\% of ranked substrates, \(n_{+}\) is the total number of positive substrates for that enzyme, and \(k = \max(1, \lfloor 0.1N \rfloor)\), with \(N\) denoting the total number of substrates tested for that enzyme. Because the maximum attainable enrichment depends on the number of tested substrates and positive substrates for a given enzyme, EF@10\% was normalized by its theoretical maximum,

\[
EF_{10\%}^{\max} = \min\left(10, \frac{N}{n_{+}}\right).
\]

The normalized enrichment factor at 10\% was then calculated as

\[
NEF_{10\%} = \frac{EF_{10\%}}{EF_{10\%}^{\max}}.
\]

Under random ranking, the expected value of EF@10\% is 1. The corresponding random baseline for the normalized enrichment factor is therefore

\[
NEF_{10\%}^{\mathrm{random}} = \frac{1}{EF_{10\%}^{\max}}.
\]

We further quantified improvement above this baseline as

\[
\Delta NEF_{10\%} = NEF_{10\%} - NEF_{10\%}^{\mathrm{random}}.
\]

This procedure was repeated for \(\Delta\)NEF@20\%. 

\subsection{Creation of the CYP database, and the CYP ranking and training datasets}
We first obtained the IDs and sequences of all manually curated enzymes from UniProt matching the CYP Pfam domain (PF00067), as well as each enzyme’s cofactors, substrates, and products, and the taxonomy of its host organism. RHEA and ChEBI were used to collect the SMILES strings of all reaction components. Non-specific reaction participants, e.g. H+ and H2O, were removed from reactions, and reactions containing non-specific SMILES strings for the substrate or product (e.g. [C*]) were removed. We further augmented this dataset with reactions from p450rdb \citep{zhang2024p450rdb}. We selected non-overlapping reactions, in which the combination of substrate, product, and catalyzing enzyme was non-redundant with our collected dataset. The taxonomic information of these enzymes were collected with the R package `rgbif`. For model training, this final dataset was filtered to remove non-monooxygenation reactions. Specifically, we selected reactions in which the only change to the substrate was the addition of one Oxygen, and potentially one to two Hydrogens. For remaining reactions, we added NADPH as a cofactor and generated atom mapping numbers from the python program rxnmapper \citep{schwaller2021extraction}. We manually verified 25 reactions to ensure validity. 

To visualize the CYP tree, we ran muscle::muscle in R with default arguments, then “iqtree3 -s cypsorg\_ma.fasta -m LG+F+R5 -nt AUTO -B 1000 --bnni” to create the phylogenetic tree. For ease of visualization, the barchart around the tree depicts the sum of the recorded substrates used by every two neighboring enzymes. All code for CYP data collection and visualization is found in \url{https://gitlab.com/cyp\_pred\_repos/cytochromesp450\_database.git}.

To identify the 45 reactions that would constitute the CYP ranking dataset, we first used MMSeqs2 \citep{steinegger2017mmseqs2} with default parameters to cluster CYP sequences at 40\% identity. Although 30\% sequence identity is standard for tasks that span multiple protein families, it may assign too many sequences into a single cluster when considering only one. We instead use 40\% identity as a standard threshold for within-CYP family placement \citep{werck2000cytochromes}. Next, we used rdkit v2025.3.5 to cluster substrates at 70\% Tanimoto similarity.
We then combined cluster assignments across enzymes and substrates, and found clusters that could be entirely devoted to the holdout ranking set. Importantly, for every enzyme in a chosen cluster, its host organism must have contained at least ten CYPs, yielding at least nine negatives to rank among. We further ensured that none of the nine highly promiscuous, drug-metabolizing human CYPs \citep{chang2024deepp450} could be present in the holdout ranking dataset. With this dataset constructed, we used the remaining data as the CYP training dataset, further splitting this into an 80/20 train/validation split through MMSeqs2 clusters. This training dataset was used to train the Boltz-MLP and RF models, and for retraining of CLIPZyme, EnzymeCAGE, EZSpecificity, and FusionESP as described below.

\subsection{Dataset creation and model training of the Boltz MLP and RF}
In order to train these models in a supervised manner, we created four negative reactions per positive reaction by randomly assigning substrates to non-catalyzing enzymes. While this strategy does not generate “hard” negatives (e.g. enzyme-substrate pairs that are similar to existing positives, yet truly yielding no products), it is more likely to avoid false negatives than utilization of similar molecules in negative generation. For each reaction, we created yaml files as inputs to Boltz (v 0.4.1), containing the protein sequence, the substrate, and heme. Protein Multiple Sequence Alignments were created with localcolabfold v. 1.5.5. We ran Boltz with default parameters. We repeated this procedure for each reaction in the holdout evaluation dataset.

The cif files containing predicted structures were used to collect residue IDs of any protein residue with at least one atom positioned less than five Ångstroms away from any atom in the substrate or heme. We then selected the representation for each of these residue-ligand atom contacts from the pair representation, the model’s internal understanding of the structure before it is predicted by the “structure prediction head”.  These representations were then averaged channel-wise, and to this final embedding we added: a one-hot embedding of the element of the atom in the substrate that is predicted to be closest to the catalytic Fe in the heme, and the predicted distance between this atom and the heme. The final embedding was input into Random Forest model and Multilayer Perceptrons to generate predictions. All code, training and evaluation datasets are found in \url{https://gitlab.com/cyp\_pred\_repos/boltzcyp}.  

\subsection{Evaluating and finetuning CLIPZyme, EnzymeCAGE, EZSpecificity, and FusionESP}
To benchmark FusionESP, we initially downloaded the best performing FusionESP model and datasets used in its training, following the directions in FusionESP’s code repository. All SMILES strings were canonicalized, and any SMILES string or protein sequence in the enzyme screening datasets with a matching entry in FusionESP’s training dataset was removed. Remaining SMILES strings were tokenized with MolFormer \citep{ross2022large} (from the HuggingFace `transformers` python package), and remaining protein sequences were embedded with ESM2-3B. Embeddings were input into the model and predictions were fed directly into AUPR, AUROC and NEF calculations. All models, data and code are available at \url{https://gitlab.com/cyp\_pred\_repos/fusionesp\_cyp\_retrain}.

We retrained FusionESP on our CYP reactions using the same training and validation datasets which were used to train the MLP. We generated embeddings for all training and validation inputs using MolFormer for canonicalized SMILES, and ESM2-3B for protein sequences. We trained a model using the same hyper parameter set and architecture as used in the FusionESP paper. As in the original paper, we chose the model with the best accuracy on the validation dataset for testing on the CYP holdout-set. Model inputs for all molecules and proteins in the holdout-set were generated in the same manner as the training data.

We used the CLIPZyme repo to create the necessary training files from our CYP dataset. Predicted protein structures were downloaded from the AlphaFold database (\url{https://alphafold.ebi.ac.uk/files/}) and sequence and per-residue embeddings were generated from ESM2-650M, replicating the original training scheme. Although we used the enzymes and substrates in the CYP training data, no negatives were explicitly created as negatives are automatically generated per-batch by the model. Any inputs with unrecoverable structures or yielding errors in processing were discarded. We used the same hyper parameters as were used to create the original best performing model, utilizing the clip\_egnn.json file in the authors’ repo. We note that we also attempted fine-tuning, in which solely the final layers of the EGNN and MPNN were unfrozen, but this led to similar performance. As in the original paper, the model weights yielding the highest CLIP quantile on the validation set was used for testing using our hold-out CYP evaluation data. For each of the 45 reactions, we generated a screening set following the instructions under “Using your own screening set” in the author’s code repository. Retrained models, data, and code are available at \url{https://gitlab.com/cyp\_pred\_repos/clipzyme-finetune}.

EZSpecificity code and model checkpoints were downloaded from the authors’ Zenodo storage repository (\url{https://zenodo.org/records/17981381}). For benchmarking across \(\alpha\)-KGs and NTLs, all SMILES strings were canonicalized, and any enzyme sequence or SMILES overlapping with the training data were removed. For all remaining enzyme-substrate pairs, as well as for the enzyme-substrate pairs in the CYP training dataset, we used Boltz-predicted cif files as inputs to EZSpecificity’s data-preprocessing pipeline. We converted all cif files to pdb using obabel, and performed minimal additional reformatting to adhere to the file format of example input files provided by the authors. We used code provided within the example.ipynb file to perform subsequent preprocessing and generate model predictions. We retrained EZSpecificity’s `best\_checkpoint.pt` using hyper parameters provided by the authors either within their paper, or within the authors' "complete-full-random-all-0-complex.yml" parameters file, adhering to the file in any discrepancy. Retrained models, data, and code are available at \url{https://gitlab.com/cyp\_pred\_repos/ezspec}. 
	
After downloading the EnzymeCAGE code (from \url{https://github.com/GENTEL-lab/EnzymeCAGE}), datasets and models were accessed according to authors' instructions. We used the Boltz-predicted CIF files of CYP complexes (inclusive of enzyme, heme, and substrate) as inputs to EnzymeCAGE’s data preprocessing code. We defined the catalytic pocket as residues less than eight Å away from any atom in the substrate. We initially created inputs for each of the 45 reactions independently, following datasets found in the authors’ “case-study/withanolides” repository. We then repeated this procedure for the CYP training data. We performed finetuning for ten epochs, according to the authors’ methods when performing domain-specific finetuning. We report the evaluation performance of all five pretrained model seeds, as well as five of our own CYP-finetuned seeds. Retrained models, data, and code is available at \url{https://gitlab.com/cyp\_pred\_repos/enzymecage-cyp}.

\subsection{Boltz Recreation of CYP complexes in the PDB}
We queried the RCSB API (accessed March 31, 2025) for entries of CYPs within our database in complex with their recorded substrate via the “protein\_id” and “substrate” fields, keeping only entries with a resolution less than 2.5 Å and a release date after June 1, 2023. This yielded three entries (8A6W, 8TDQ, and 8X3E), for which we generated MSAs and inputs to Boltz as described above. We include Boltz’s structural predictions for these three complexes in the Supplementary Materials, as well as PyMol commands to quantify RMSD for the global complex, RMSD, and six substrate recognition sites (SRSs). The residues in the SRSs were determined using sequence alignment to known SRSs, the prevailing method in the literature \citep{bouille2025lineage}. Scripts for recreation can be found at \url{https://gitlab.com/cyp\_pred\_repos/boltzcyp}.

\section*{Acknowledgements}
We would like to thank the following members and collaborators of the Pluskal and Barzilay labs for helpful conversations, feedback, and advice: Alexandre Bouillé, Anton and Roman Bushuiev, Safa Mert Akmese, Matouš Soldát, Prof. Josef Sivic, Itamar Chinn, Weian Mao, Peter Mikhael, Timothy O'Donnell, and Tally Portnoi. We also thank Prof. Gaurav Moghe for thoughtful comments on the manuscript. We thank the good people of the Penn Advanced Research Computing Center, and The Infrastructure Group for computational resources and support.

\section*{Code and Data Availability}

Code and data for enzyme screen processing can be found at \url{https://github.com/lizmahood/enzyme-datasets}. Code and data for producing the CYP database can be found at \url{https://gitlab.com/cyp_pred_repos/cytochromesp450_database}. 

The following repositories contain the remainder of the code for results reproduction:

\begin{itemize}
	\item \textbf{FusionESP}: \url{https://gitlab.com/cyp_pred_repos/fusionesp_cyp_retrain}
	\item \textbf{EZSpecificity}: \url{https://gitlab.com/cyp_pred_repos/ezspec}
	\item \textbf{CLIPZyme}: \url{https://gitlab.com/cyp_pred_repos/clipzyme-finetune}
    \item \textbf{EnzymeCAGE}: \url{https://gitlab.com/cyp_pred_repos/enzymecage-cyp}
    \item \textbf{BoltzCYP}: \url{https://gitlab.com/cyp_pred_repos/boltzcyp}
\end{itemize}

All datasets and models can be found at \url{https://huggingface.co/datasets/lizmahood/cyp_pred_repos/tree/main}. 

\bibliographystyle{vancouver}
\bibliography{citation}

\clearpage
\setcounter{figure}{0}
\renewcommand{\thefigure}{S\arabic{figure}}
\setcounter{table}{0}
\renewcommand{\thetable}{S\arabic{table}}

\begin{center}
    {\LARGE\bfseries\color{PennBlue} Supplementary Information}
\end{center}
\vspace{1em}

\begin{figure}[H]
    \centering
    \includegraphics[width=1\textwidth]{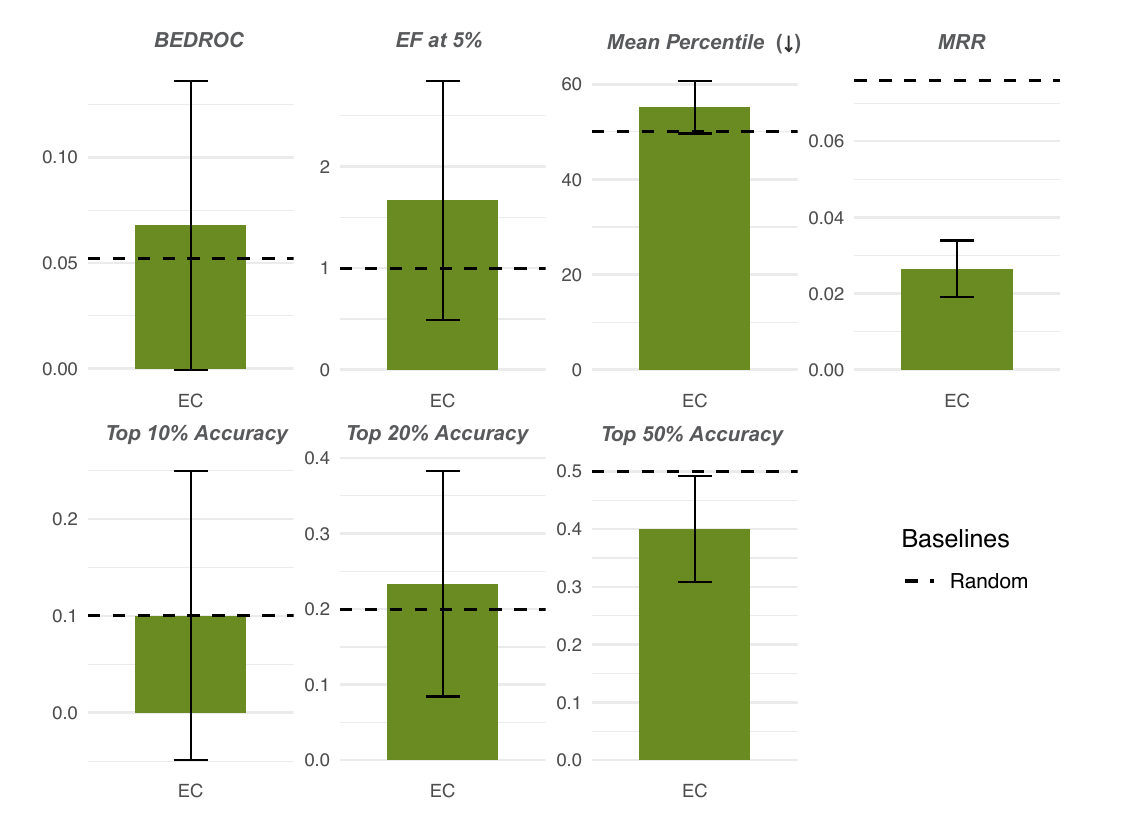}
    \caption{\textbf{Performance of EnzymeCAGE on a leakage-resolved CYP ranking dataset.} To create this “leakage-resolved” dataset, any enzyme or substrate that appeared in EnzymeCAGE’s training data was removed from our CYP ranking dataset. These plots show EnzymeCAGE’s performance across the remaining six (down from 45) reactions. Five model seeds were evaluated. The mean score for each metric is plotted as the bar height, error bars represent standard deviation. Higher scores are better for all metrics excepting Mean Percentile. BEDROC = Boltzmann-Enhanced Discrimination of the Receiver Operating Characteristic ($\alpha$ at 20); EF = Enrichment Factor; MRR = Mean Reciprocal Rank.}
    \label{fig:S_EC_pruned}
\end{figure}

\begin{figure}[H]
    \centering
    \includegraphics[width=1\textwidth]{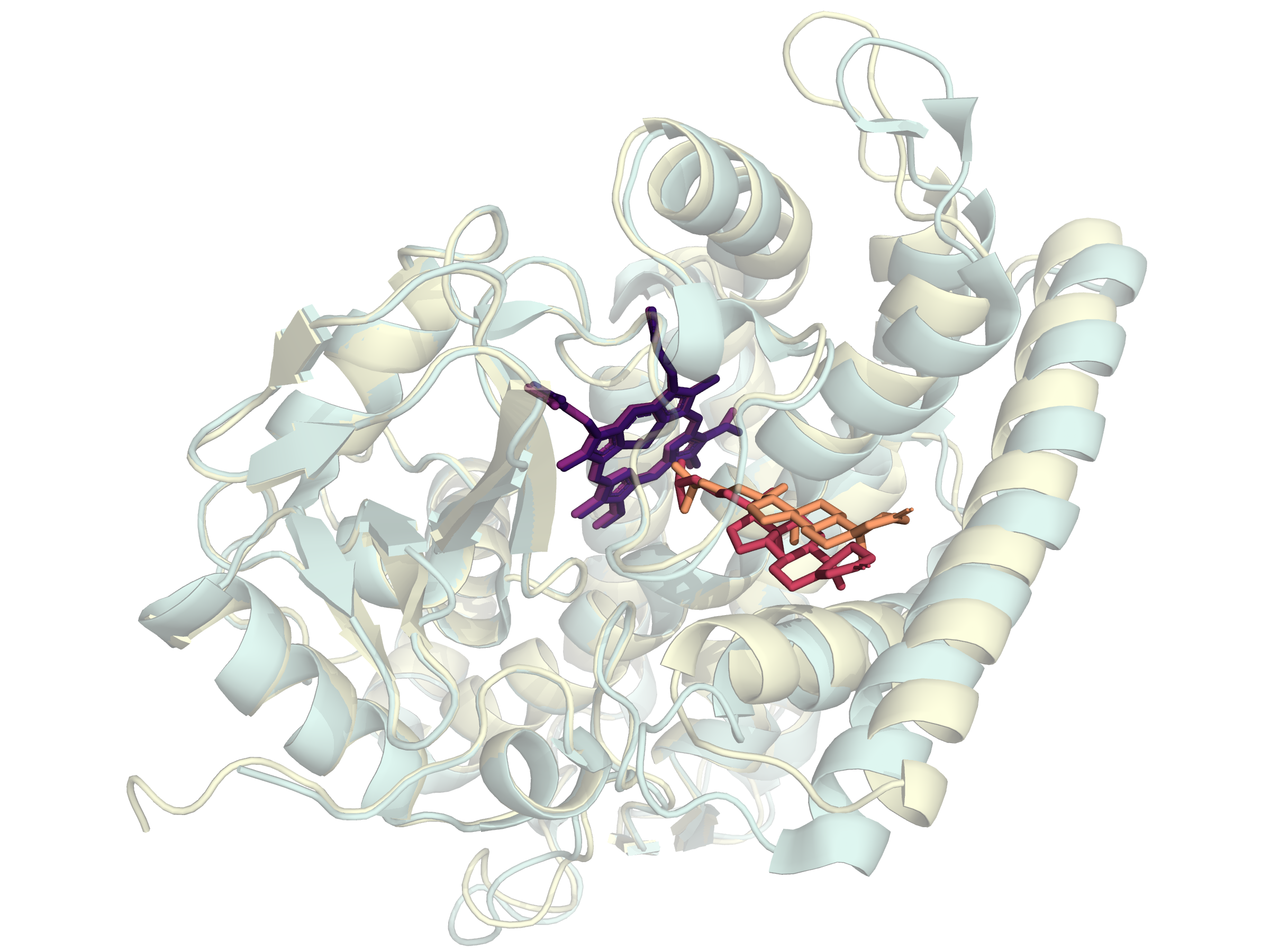}
    \caption{\textbf{Global alignment between Boltz-predicted and actual structures for PDB ID 8a6w.} Colors are as in Figure \ref{fig:F4-boltz}: Wheat, Navy and Orange represented the predicted complex, and Mint, Purple, and Magenta represent the true crystal structure. 8a6w is CYP142 from \textit{Mycobacterium tuberculosis} in complex with cholestenone.}
    \label{fig:S_8a6w}
\end{figure}

\begin{figure}[H]
    \centering
    \includegraphics[width=1\textwidth]{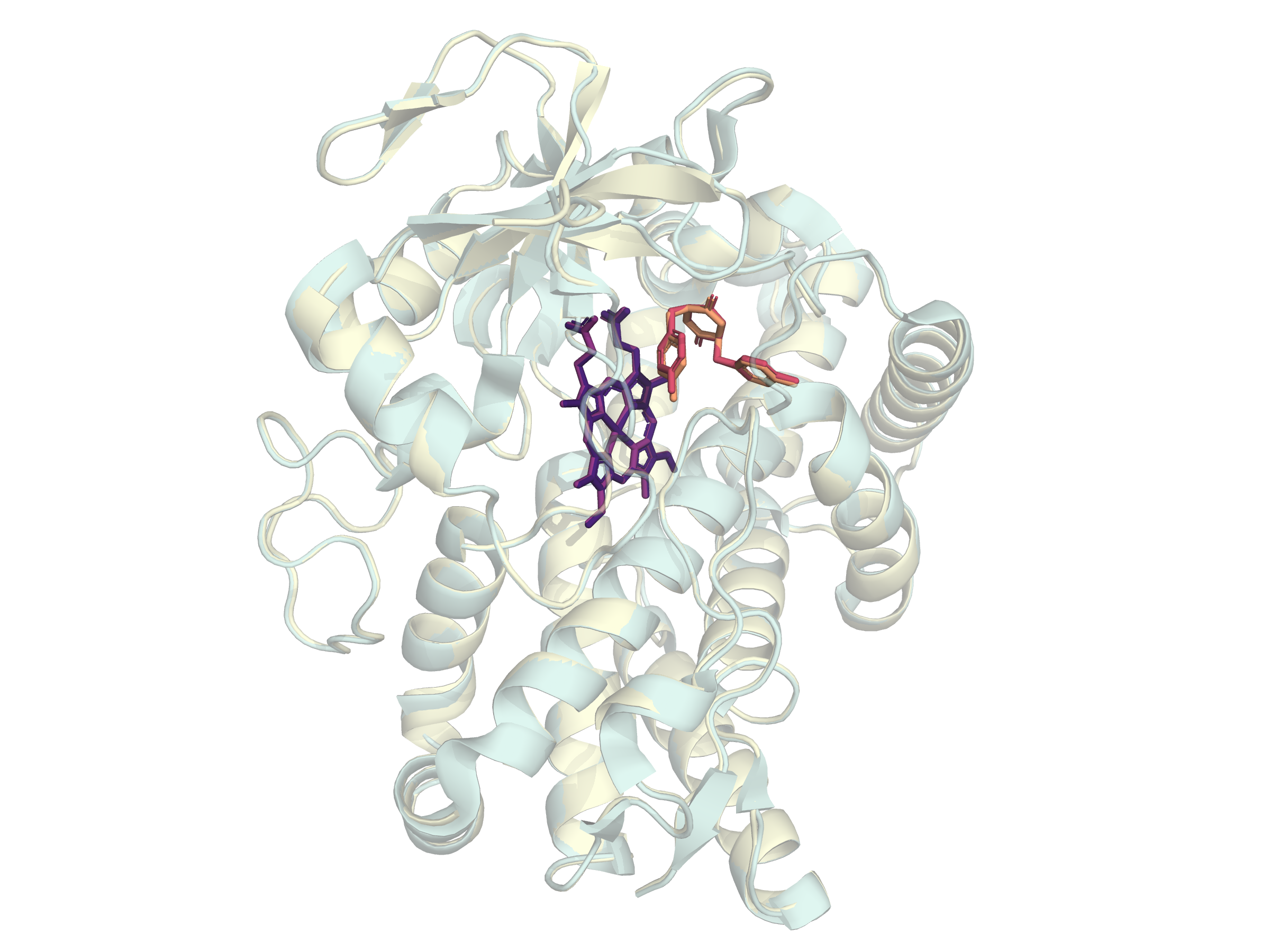}
    \caption{\textbf{Global alignment between Boltz-predicted and actual structures for PDB ID 8tdq.} Colors are as in Figure \ref{fig:F4-boltz}: Wheat, Navy and Orange represented the predicted complex, and Mint, Purple, and Magenta represent the true crystal structure. 8tdq is CYP121 from \textit{Mycobacterium tuberculosis} in complex with \textit{cyclo}(l-tyrosine-l-tyrosine).}
    \label{fig:S_8tdq}
\end{figure}

\begin{figure}[H]
    \centering
    \includegraphics[width=1\textwidth]{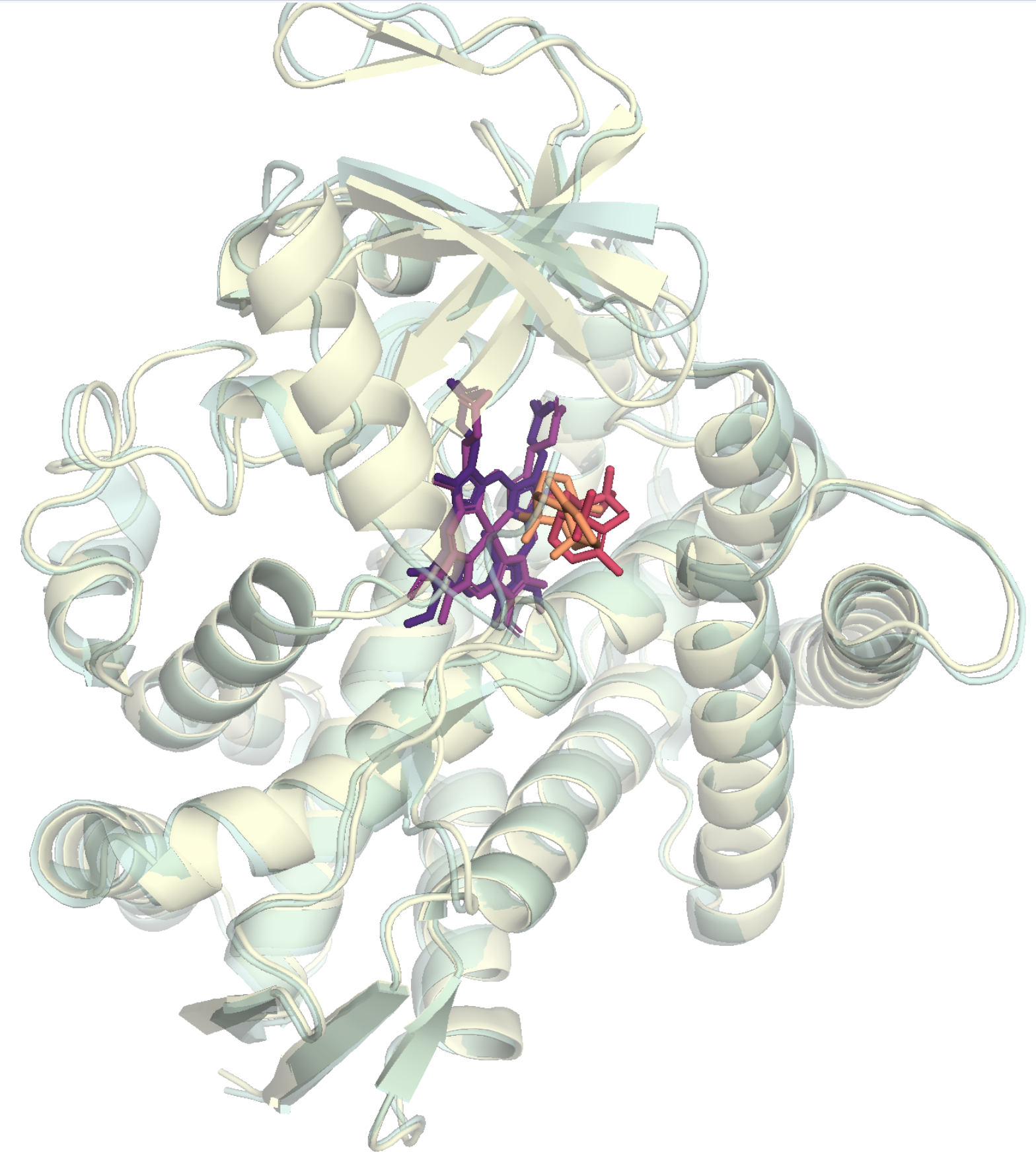}
    \caption{\textbf{Global alignment between Boltz-predicted and actual structures for PDB ID 8x3e.} Colors are as in Figure \ref{fig:F4-boltz}: Wheat, Navy and Orange represented the predicted complex, and Mint, Purple, and Magenta represent the true crystal structure. 8x3e is CYP725A4 from \textit{Taxus cuspidata} in complex with taxadiene.}
    \label{fig:S_8x3e}
\end{figure}

\begin{figure}[H]
    \centering
    \includegraphics[width=1\textwidth]{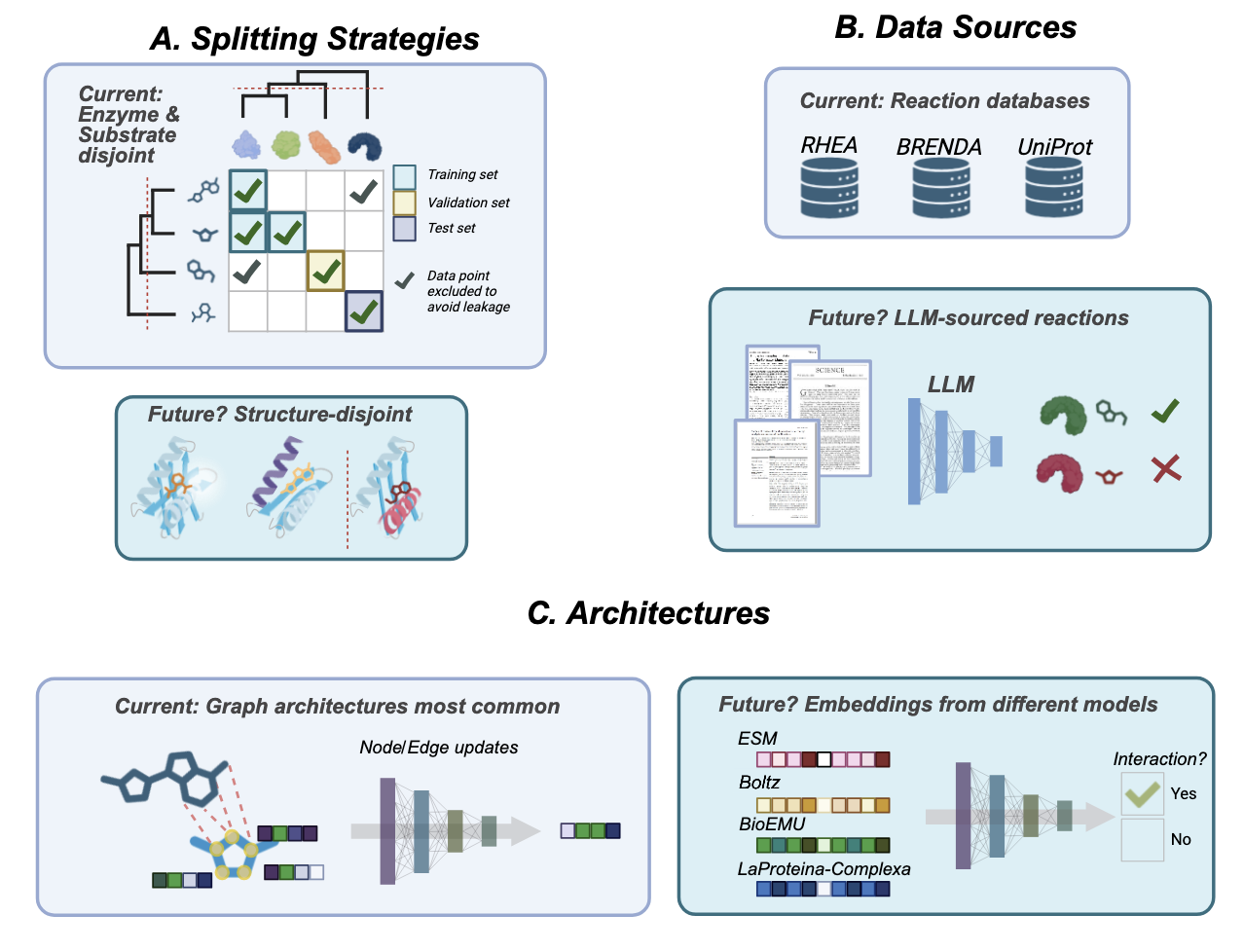}
    \caption{\textbf{Potential future directions of enzyme-substrate prediction models} \textbf{(A)} Current splitting strategies consider enzymes and substrates separately. In the protein-lignad domain, such splits can create leakage that structure-aware splits have ameliorated \citep{bushuiev2024revealing}. \textbf{(B)} Current models source data from reaction databases, which lag behind the primary literature and do not contain negatives. While negatives can currently be drawn from enzyme screens, these are limited in number. In the future, we hope function-abolishing site-directed mutagenesis experiments can be incorporated as a source of additional negatives, perhaps drawn from primary literature using LLMs as has been done for positive enzyme-substrate reactions \citep{smith2025funcfetch}. \textbf{(C)} Since movement is a critical component of substrate entry and catalysis in many enzyme families, we believe that deriving enzyme representations from conformation-aware models, including LaProteina-Complexa \citep{didi2026scaling} or BioEMU \citep{lewis2025scalable}} 
    \label{fig:S_future}
\end{figure}

\begin{threeparttable}[ht]
\centering
\scriptsize
\setlength{\tabcolsep}{4pt}
\caption{\textbf{Substrate prediction model benchmark results by model and enzyme class.} ma = macro; mi = micro; Rand. = random baseline; Prev. = prevalence; Enz. = enzymes; Sub. =  substrates; EZSp = EZSpecificity; FESP = FusionESP; NTL = Nitrilases; \(\alpha\)-KG = \(\alpha\)-ketoglutarate/Fe(ii)-dependent enzymes.}
\label{tab:benchmark_results}
\begin{tabular}{llrrrrrrrrrrrr}
\toprule
 &  & \multicolumn{4}{c}{\textbf{Performance} }
    & \multicolumn{2}{c}{\textbf{Macro AUPR enrichment}}
    & \multicolumn{4}{c}{\textbf{Dataset}}
    & \multicolumn{2}{c}{\textbf{Recall}} \\
\cmidrule(lr){3-6}
\cmidrule(lr){7-8}
\cmidrule(lr){9-12}
\cmidrule(lr){13-14}
\textbf{Model} & \textbf{Class} 
& maAUROC & maAUPR & miAUROC & miAUPR 
& $>$ Rand. & $>$ 2$\times$ Rand.
& n Enz. & n Sub. & n Pairs & Prev.
& @10 & @20 \\
\midrule
EZSp & NTL  & 0.743 & 0.208 & 0.556 & 0.134 & 40.260 & 12.987 & 77  & 10 & 755  & 0.155 & 0.068 & 0.154 \\
EZSp & \(\alpha\)-KG & 0.675 & 0.073 & 0.551 & 0.049 & 53.333 & 22.667 & 75  & 66 & 4608 & 0.046 & 0.081 & 0.156 \\
FESP & NTL  & 0.671 & 0.175 & 0.532 & 0.118 & 60.504 & 21.849 & 119 & 24 & 2796 & 0.105 & 0.099 & 0.215 \\
FESP & \(\alpha\)-KG & 0.700 & 0.094 & 0.559 & 0.054 & 54.430 & 27.848 & 79  & 71 & 5248 & 0.041 & 0.149 & 0.228 \\
\bottomrule
\end{tabular}
\scriptsize
\end{threeparttable}

\begin{table}[ht]
\centering
\small
\setlength{\tabcolsep}{4pt}
\begin{tabular}{llrrrrrrr}
\toprule
\textbf{Model} & \textbf{Class} & \textbf{$k$} & \textbf{n Enzymes} & \textbf{n $>$ Rand. \(\Delta\)NEF} & \textbf{\% $>$ Rand. \(\Delta\)NEF} & \textbf{Median \(\Delta\)NEF} & \textbf{q25 \(\Delta\)NEF} & q75 \textbf{\(\Delta\)NEF} \\
\midrule
EZSp & NTL  & 10\% & 77  & 8  & 10.390 & -0.100 & -0.200 & -0.100 \\
EZSp & NTL  & 20\% & 77  & 15 & 19.481 & -0.200 & -0.200 & -0.200 \\
EZSp & \(\alpha\)-KG & 10\% & 75  & 12 & 16.000 & -0.100 & -0.100 & -0.100 \\
EZSp & \(\alpha\)-KG & 20\% & 75  & 24 & 32.000 & -0.200 & -0.200 &  0.133 \\
\midrule
FESP     & NTL  & 10\% & 119 & 22 & 18.487 & -0.100 & -0.106 & -0.100 \\
FESP     & NTL  & 20\% & 119 & 38 & 31.933 & -0.200 & -0.200 &  0.129 \\
FESP     & \(\alpha\)-KG & 10\% & 79  & 18 & 22.785 & -0.100 & -0.100 & -0.100 \\
FESP     & \(\alpha\)-KG & 20\% & 79  & 22 & 27.848 & -0.200 & -0.200 &  0.092 \\
\bottomrule
\end{tabular}
\caption{\textbf{\(\Delta\)NEF results by model, screening set, and retrieval threshold.} EZSp = EZSpecificity; FESP = FusionESP; NTL = Nitrilases; \(\alpha\)-KG = \(\alpha\)-ketoglutarate/Fe(ii)-dependent enzymes}
\label{tab:dnef_results}
\end{table}

\begin{table*}[ht]
\centering
\caption{\textbf{RMSD of Boltz structural predictions.} Complexes were pulled from the PDB if they contained a cytochrome p450 in complex with its native substrate and were deposited after June 1, 2023. Following prior studies, Substrate Recognition Sites (SRS) were calculated by sequence alignment of each CYP to CYPs with known SRSs. All alignments were performed in PyMol (see Supplemental for commands).}
\label{tab:S_boltz_rmsd}
\small
\setlength{\tabcolsep}{6pt}
\begin{tabular}{lccccccccc}
\toprule
\textbf{PDB Complex} & \textbf{Global alignment} & \textbf{CA alignment} & \textbf{heme} & \textbf{SRS1} & \textbf{SRS2} & \textbf{SRS3} & \textbf{SRS4} & \textbf{SRS5} & \textbf{SRS6} \\
\midrule
8TDQ & 0.227 & 0.214 & 0.035 & 0.128 & 0.060 & 0.051 & 0.082 & 0.087 & 0.070 \\
8A6W & 0.408 & 0.359 & 0.051 & 0.201 & 1.269 & 0.085 & 0.539 & 0.163 & 0.193 \\
8X3E & 0.608 & 0.566 & 0.174 & 3.401 & 2.114 & 0.148 & 0.413 & 3.405 & 3.546 \\
\bottomrule
\end{tabular}
\end{table*}

\end{document}